\documentclass[conference, a4paper]{IEEEtran}
\IEEEoverridecommandlockouts
\usepackage{amsmath,amssymb,amsfonts}
\usepackage{algorithmic}
\usepackage{graphicx}
\usepackage{textcomp}
\usepackage{xcolor}
\usepackage{tabularx}
\usepackage{cite}

\def\BibTeX{{\rm B\kern-.05em{\sc i\kern-.025em b}\kern-.08em
    T\kern-.1667em\lower.7ex\hbox{E}\kern-.125emX}}
\begin{document}

\title{LEMO: Learn to Equalize for MIMO-OFDM Systems with Low-Resolution ADCs\\
\thanks{Dr. Chu's  work  is partially supported by the SJTU overseas study grant and by N.S.F. of China under Grant No.61871265. Dr. Pei 's  work  is supported by the National Nature Science Foundation of China (NSFC) under Grant 61873163 and Grant 61402283. Dr. Qiu's  work  is partially supported  by  NSFC under Grant No.61571296. }
}

\author{
\IEEEauthorblockN{Lei Chu\IEEEauthorrefmark{1}\IEEEauthorrefmark{2}, Ling Pei\IEEEauthorrefmark{1}, 
Husheng Li\IEEEauthorrefmark{2}, Robert Caiming Qiu\IEEEauthorrefmark{1}\IEEEauthorrefmark{3}\IEEEauthorrefmark{4}}
\IEEEauthorblockA{\IEEEauthorrefmark{1}School of Electronics, Information and Electrical Engineering, Shanghai Jiao Tong University, Shanghai, China}
\IEEEauthorblockA{\IEEEauthorrefmark{2}Department of Electrical Engineering and Computer Science, University of Tennessee, Knoxville, USA}
\IEEEauthorblockA{\IEEEauthorrefmark{3}Department of Electrical and Computer Engineering, Tennessee Technological University, Cookeville, USA \\
\IEEEauthorblockA{\IEEEauthorrefmark{4}School of Electronic Information and Communications, Huazhong University of Science and Technology, China}
(leochu@sjtu.edu.cn; ling.pei@sjtu.edu.cn; \ hli31@utk.edu; \ caiming@hust.edu.cn)}
}

\maketitle

\begin{abstract}
		
This paper develops a new deep neural network optimized equalization framework for massive multiple input multiple output
orthogonal frequency division multiplexing (MIMO-OFDM) systems that employ low-resolution analog-to-digital converters (ADCs) at the base
station (BS). The use of low-resolution ADCs could largely reduce hardware complexity and circuit power consumption, however, it makes the
channel station information almost $blind$ to the BS, hence causing difficulty in solving the equalization problem.
In this paper, we consider a supervised learning architecture, where the goal is to learn a representative function that can predict the
targets (constellation points) from the inputs (outputs of the low-resolution ADCs) based on the labeled training data (pilot signals).
Especially, our main contributions are two-fold: 1) First, we design a new activation function, whose outputs are close to the constellation
points when the parameters are finally optimized, to help us fully exploit the stochastic gradient descent method for the discrete
optimization problem. 2) Second, an unsupervised loss is designed and then added to the optimization objective, aiming to enhance
the representation ability (so-called generalization). Lastly, various experimental results confirm the superiority of the proposed equalizer over some existing ones, particularly when the statistics of the channel state information are unclear.

\end{abstract}


\section{Introduction}

Massive MIMO is foreseen to be one of the key enablers for the future generation communication system,
in which the spectral efficiency is expected to be several-order higher than that in the current one
\cite{Lu2014An}.  Scaling up the number of the antennas in the BS can offer numerous advantages \cite{Rusek2012Scaling} while
exhibiting the gigantic energy consumptions. To alleviate this predicament, equipping the BS with low-resolution ADCs/DACs can greatly reduce
energy consumptions \cite{Wang2015Multiuser,Wang2016Signal, Wang2017Quantization, Chu2019E} in the massive MIMO system.

Last few years have witnessed the publication of many technical results reporting the performance of the massive MIMO system
with low-resolution ADCs. In this work, we will focus on the data equalization problem. For the frequency-flat channel, the authors of
\cite{Jacobsson2017Throughput} show that the performance gap between the quantized MIMO system and ideal one can be approached by increasing
the number of antennas. Similar results have been reported in the frequency-selective case, i.e.,
single-cell sparse broadband massive MIMO \cite{Mezghani2017Blind} and then multi-cell millimeter wave massive MIMO \cite{Xu2018Performance}.

Most of these works are based on the Bussgang theorem \cite{Julian1952Crosscorrelation}, with which the severe nonlinearity introduced by the ADCs is approximately expressed by a linear combination of the input and a distortion term. The Bussgang theorem based data equalization methods have
low computational complexity comparable to the classical linear data equalization methods. On the other hand, they have
to suffer from the performance saturation pain in the mid-to-high SNR regime \cite{jeon2018one, myers2019message, Wen2016Bayes}and attest to a low
degree of adaptability in practical 5G millimeter-wave massive MIMO systems \cite{Zhang2018On}.


It is hard to understand the nonlinearity introduced by the ADCs with a tractable mathematical model.
 As a result, the best approach for data equalization problem is unclear. Alternatively, it would be interesting to
comprehend (or learn) the nonlinear structure of the employed ADCs leveraging the potential of the
deep learning based methods \cite{Lecun2015Deep}.

Recently, the deep learning based detection methods have been proposed for MIMO-OFDM systems with ideal DACs \cite{Ye2018Power, 8646357}.
It has been shown that deep learning based method has comparable performance with the minimum
mean-square error receiver and shows robustness in the case of nonlinear clipping noise. The authors of \cite{Farsad2018Neural} use
a recurrent neural network based approach to detect data sequences in molecular communication systems with blind channel state information.
Besides, supervised-learning-aided estimator has been proposed for the frequency-flat MIMO system with low-resolution ADCs
\cite{Jeon2016Supervised}.

In this paper, instead of relying solely on deep learning methods, we propose a new deep neural network optimized equalization framework,
jointly exploiting structural knowledge from MIMO systems and harnessing the power of unsupervised deep learning,
for MIMO-OFDM systems with low-resolution ADCs.

The remainder of this paper is structured as follows. Section \ref{Sec:2} introduces the signal model for
the MIMO-OFDM systems with low-resolution ADCs, in which the Bussgang theorem based linearized data equalization method is discussed.
In Section \ref{Sec:3}, we detail the design of the proposed method. A coarse deep neural network based equalizer,
following the structure of the widely-used supervised learning method, is firstly proposed. To enhance the generalization of proposed equalizer,
a fine deep neural network based equalizer is then developed by leveraging the knowledge from the unsupervised learning. Furthermore,
numerical case studies in Section \ref{Sec:4} are provided to evaluate the performance of the proposed approach.
Lastly, the conclusion and acknowledgment of this paper are given in Section \ref{Sec:5} and Section \ref{Sec:6}, respectively.

$Notations:$ Throughout this paper, vectors and matrices are given in
lowercase and uppercase boldface letters, e.g., $\bf{x}$ and $\bf{X}$, respectively.
We use ${{\bf{X}} ^{H}}$ to denote the conjugate transpose of ${\bf{X}}$.
The $k$th row and $l$th column element of ${\bf{X}}$ is denoted by ${\left[ {\bf{X}} \right]_{kl}}$.
We use $\Re \left( {\bf{x}} \right)$, $\Im \left( {\bf{x}} \right)$,
and ${\left\| {\bf{x}} \right\|_2}$ to represent the real part, the imaginary part,
$\ell_2$-norm of vector ${\bf{x}}$.

\section{PRELIMINARIES}
\label{Sec:2}

\subsection{MIMO-OFDM System Model}

We consider a MIMO-OFDM uplink system \cite{Goldsmith2005Wireless} with $R$ receive antennas at the BS, serving simultaneously $K$
single-antenna user terminals (UTs) over $N$ subcarriers.  We use ${\bf{h}}_{rk} \in {\mathbb{C}^{\mu}}$ to denote the channel impulse response
of $\mu$ taps between the $r$-th receive antenna and the $k$-th transmit antenna, for $k = 1,\cdots,K$, $r = 1,\cdots,R$.
Let ${\bf{X}}_{k} \in {\mathbb{C}^{N \times T}}$
be a sequence of $T$ OFDM symbols to be sent from the $k$-th antenna. After removing the cyclic prefix and applying an IFFT,
 the received signal ${{\bf{Y}}_r} \in {\mathbb{C}^{N \times T}}$ at the $r$-th receive antenna reads
\begin{equation}
\label{eq1}
{{\bf{Y}}_r} = \sum\limits_{k = 1}^K {{{\bf{H}}_{rk}}{{\bf{F}}^H}{{\bf{X}}_k} + {{\bf{V}}_r}},
\end{equation}
where ${{\bf{H}}_{rk}}$ denotes the circulant channel convolution matrix, ${\bf F}$ is the $unitary$ discrete Fourier transform (DFT)
 matrix \cite{Goldsmith2005Wireless}  of dimension $N \times N$  and the elements of ${{\bf{V}}_r}$ are assumed to be identically
independent distributed zero-mean Gaussian variables with variance $\sigma^2$.

\subsection{Quantization}

For the MIMO-OFDM uplink system with low-resolution ADCs, it is assumed that the real and imaginary parts of received signals are
quantized separately by a $B$-bit symmetric uniform quantizer $Q$, denoted by
\begin{equation}
\label{eq2}
{\bf{z}} = Q({\bf{y}}) = {Q( \Re \left( {{\bf{y}}} \right))} + {j Q( \Im\left( {{\bf{y}}} \right))}.
\end{equation}
Specially, the real-valued quantizer $Q(.)$ maps the input to a set of labels
$\Omega = \left\{ {{l_0},{\kern 1pt} \cdots ,{l_{{2^B-1}}}} \right\}$, which are determined by the set of thresholds
 $\Gamma  = \left\{ {{\tau _0},{\kern 1pt}  \cdots ,{\tau _{{2^B}}}} \right\}$, such
 that $ - \infty  = {\tau _0} <  \cdots  < {\tau _{{2^B}}} = \infty $. For a $B$-bit ADC with step size $\Delta$, the thresholds and
 quantization labels are respectively given by \cite{Jacobsson2017Throughput, Chu2019R}
\begin{equation}
\label{eq3}
{l_b} = \Delta \left( {b - \frac{{{2^{B}- 1}}}{2}} \right), \  b = 0, \cdots ,{2^B -1},
\end{equation}
and
\begin{equation}
\label{eq4}
{\tau _b} = \Delta \left( {i - {2^{B - 1}}} \right), \ b = 1, \cdots ,{2^B} - 1 .
\end{equation}
In the case of 1 bit ADCs, the output set reduces to ${\Omega_1} = \left\{ { \pm \frac{\Delta }{2} \pm {\rm i} \frac{\Delta }{2}} \right\}$.

From \eqref{eq1} and  \eqref{eq2}, the resulting quantized signals yield
\begin{equation}
\label{eq5}
{{\bf{Z}}_r} = Q\left( {{{\bf{Y}}_r}} \right) = Q\left( {\sum\limits_{k = 1}^K {{{\bf{H}}_{rk}}{{\bf{F}}^H}{{\bf{X}}_k} + {{\bf{V}}_r}} } \right).
\end{equation}

\subsection{Data Equalization and Linearized Solution}

The serve nonlinearity of \eqref{eq2} makes the channel station information almost $blind$ to the BS, resulting in a challenging work for data
equalization in quantized frequency-selective MIMO systems. Recently, as shown in the Section I, many researchers proposed to represent
the nonlinearity of a Gaussian signal by the sum of a linear transformation and an uncorrelated quantization error term, according to the
Bussgang theorem \cite{Julian1952Crosscorrelation}. Following the Bussgang theorem, \eqref{eq5} can be approximately represented by

\begin{align}
\label{eq6}
{{\bf{Z}}_r} &= Q\left( {\sum\limits_{k = 1}^K {{{\bf{H}}_{rk}}{{\bf{F}}^H}{{\bf{X}}_k} + {{\bf{V}}_r}} } \right)  \notag \\
& = {\bf{G}}\left( {\sum\limits_{k = 1}^K {{{\bf{H}}_{rk}}{{\bf{F}}^H}{{\bf{X}}_k} + {{\bf{V}}_r}} } \right) + {\bf{W}}_r  \notag  \\
 &= {\bf{G}}\sum\limits_{k = 1}^K {{{\bf{H}}_{rk}}{{\bf{F}}^H}{{\bf{X}}_k}}  + {\bf{D}}_r,
\end{align}
where ${\bf{G}}$ is a diagonal matrix and ${\bf{D}}_r$ is referred to the quantization error. The circulant channel matrix
has an eigenvalue decomposition
\begin{equation}
\label{eq7}
{{\bf{H}}_{{r}{k}}} = {{\bf{F}}^{{H}}}{{\bf{\Lambda }}_{{r}{k}}}{\bf{F}},
\end{equation}
where ${{\bf{\Lambda }}_{{r}{k}}}$ is a diagonal matrix and ${{\bf{\Lambda }}_{{r}{k}}} = {\mathop{\rm diag}\nolimits}
\left\{ {\sqrt N {\bf{F}}_{N \times \mu }^{{H}}{{\bf{h}}_{{r}{k}}}} \right\} $. Substituting  \eqref{eq7} into \eqref{eq6} can give the
subcarrier-wise input-output relationship:
\begin{equation}
\label{eq8}
{{\bf{Z}}_n} = \rho {{\bf{H}}_{n}}{{\bf{X}}_n} + {{{\bf{\tilde D}}}_n},
\end{equation}
where $\rho$ is the Bussgang decomposition factor. Then the Bussgang theorem based data equalization problem can be formulated as
\begin{equation}
\label{eq9}
\begin{array}{*{20}{c}}
{\min }&{\sum\limits_{n = 1}^N {\left\| {{{\bf{Z}}_n} - \alpha {{\bf{H}}_n}{{\bf{X}}_n}} \right\|_2^2} }\\
{{\mathop{\rm s.t.}\nolimits} }&{{{\left[ {{{\bf{X}}_u}} \right]}_{ij}} \in \Omega }
\end{array},
\end{equation}
where $\Omega$ is the set of the constellation points. Discarding the nonconvex constraint
${{{\left[ {{{\bf{X}}_u}} \right]}_{ij}} \in \Omega }$,
one can  use the minimum mean squared error (MMSE) based method \cite{Poor1994An} for the channel estimation and then data equalization.
Unluckily, such a linearized
data equalization method has to suffer from the performance saturation in the mid-to-high SNR regime. Specially, the BER curves achievable
with linearized data equalization methods saturate at certain finite SNR, above which no further improvement can be obtained. As a result, it is
vital to understanding the nonlinearity of \eqref{eq2} by the new techniques, i.e., the deep neural networks based methods.

\section{LEMO: Learn to Equalize for MIMO-OFDM}
\label{Sec:3}

\begin{figure}[htbp]
\centerline{\includegraphics[width=0.475\textwidth]{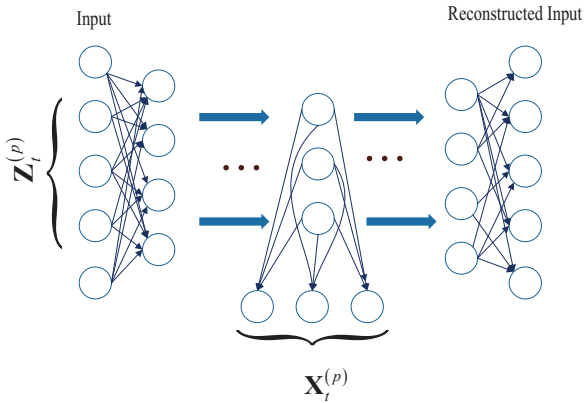}}
\caption{The MIMO-OFDM system with low-resolution ADCs and the proposed deep neural network based equalizer.}
\label{fig1}
\end{figure}

For the coarsely quantized MIMO-OFDM system, the true channel state information is almost completely unknown, the best method to data equalization
becomes unclear. As shown in Fig. \ref{fig1}, we propose a deep learning based method to train a data-driven detector to determine transmitted symbols
from pilots.

\subsection{Coarse Deep Neural Network based Equalizer}

We start by applying the supervised deep learning \cite{Lecun2015Deep} for data equalization in quantized MIMO-OFDM system.
The employed method in subsection is referred to the coarse deep neural network (CDNN) based equalizer.

\subsubsection{Data Preprocessing}
Generally, the supervised deep learning
based approach has two parts: Offline training and online test. Let $T$ be the number of collected data samples. For $t=1,\cdots,T$, we use
\[\left\{ {{{\bf{Z}}_t},{{\bf{X}}_t}} \right\} = \left[ {\begin{array}{*{20}{c}}
{\underbrace {\left\{ {{\bf{Z}}_t^{\left( p \right)},{\bf{X}}_t^{\left( p \right)}} \right\}}_{{\rm{pilot \ singal}}}}&
{\underbrace {\left\{ {{\bf{Z}}_t^{\left( t \right)},{\bf{X}}_t^{\left( t \right)}} \right\}}_{{\rm{test \ signal}}}}
\end{array}} \right]\]
to represent data collected at $t$-th sampling.
In this paper, the elements of ${{\bf{Z}}_t}$ and ${{\bf{X}}_t}$ are the subcarrier-wise real-valued received signals
(outputs of low-resolution DACs) and the transmitted signals, respectively.

\subsubsection{Network Parameters Optimization}

At the offline training stage, the pilot data set:
\[\left\{ {{{\bf{Z}}^{\left( p \right)}},{{\bf{X}}^{\left( p \right)}}} \right\} = \left\{ {\left( {{\bf{Z}}_1^{\left( p \right)},
{\bf{X}}_1^{\left( p \right)}} \right), \cdots ,\left( {{\bf{Z}}_T^{\left( p \right)},{\bf{X}}_T^{\left( p \right)}} \right)} \right\}\]
is used to find the optimal set of parameters of a $U$-layer neural network by solving the optimization problem:
\begin{equation}
\label{eq31}
{\cal L}\left( \Theta_0  \right) = \mathop {{\rm{argmin}}}\limits_{\Theta_0}  \left\| {{{\bf{Z}}^{\left( p \right)}} - {\varphi _S}\left( {{{\bf{W}}_S}{{{\bf{\tilde X}}}^{\left( p \right)}}} \right)} \right\|_2^2,
\end{equation}
where \[{{{\bf{\tilde X}}}^{\left( p \right)}} = {\varphi _U}\left( {{{\bf{W}}_U} \cdots {\varphi _1}\left( {{{\bf{W}}_1}{{\bf{X}}^{\left( p \right)}}} \right)} \right),\] 
for $u=1,\cdots, U$, $\varphi_u$ is the activation function and ${\bf{W}}_u$ are weight matrix at $u$-th layer. $\varphi_S$ is a specially designed activation which will be explained latter.
The parameters ${\Theta_0 = \left\{ {{{\bf{W}}_1}, \cdots ,{{\bf{W}}_U}, {{\bf{W}}_S}} \right\}}$ are updated  to minimize the expected
loss \eqref{eq31} by using a stochastic descent based method \cite{Lecun2015Deep}, i.e., stochastic gradient decent (SGD), or Adam.

The deep learning based equalizer described above tries to generate a nonlinear relationship between the transmitted signals and the
quantized received signals according to the supervised task,
hopefully aiming at understanding (or learning) the true channel matrix information that has been almost $destroyed$ by
the use of low-resolution ADCs.

However, solely training a supervised neural network is likely to handle a problem with under-constrained configurations,
and will find a solution that can well fit the training data but can not generalize well \cite{Zhang17U}, especially in the case of
neural discrete representation learning (the data equalization problem is itself a discrete optimization problem
\cite{Poor1994An}).
These limitations motivate us to design a fine deep neural network (FDNN) based equalizer .

\subsection{Fine Deep Neural Network based Equalizer}

Developing a deep neural network with high generalization ability plays a key role for the data equalization of the quantized frequency-selective
MIMO systems. Generally, increasing the number of the layers and the number of neurons in a layer can help improve the generalization ability of
deep neural networks \cite{Lecun2015Deep}. However, training a very deep neural network has to face up with the vanishing or exploding
gradient problem \cite{NIPS2018_7339}. Besides, the real time requirement prevents the widespread use of the very deep neural network.

\subsubsection{The Skew-symmetric Weights Matrix}

We stat with a simple example. Let ${\bf x}$ and ${\bf y}$ be two complex-valued vectors and ${\bf H}$ a complex-valued matrix. The equation
\[{\bf{y = Hx}}\]
equals to
\[\left[ {\begin{array}{*{20}{c}}
{{\mathop{\rm Re}\nolimits} \left\{ {\bf{y}} \right\}}\\
{{\mathop{\rm Im}\nolimits} \left\{ {\bf{y}} \right\}}
\end{array}} \right] = \left[ {\begin{array}{*{20}{c}}
{{\mathop{\rm Re}\nolimits} \left\{ {\bf{H}} \right\}}&{ - {\mathop{\rm Im}\nolimits} \left\{ {\bf{H}} \right\}}\\
{ {\mathop{\rm Im}\nolimits} \left\{ {\bf{H}} \right\}}&{{\mathop{\rm Re}\nolimits} \left\{ {\bf{H}} \right\}}
\end{array}} \right]\left[ {\begin{array}{*{20}{c}}
{{\mathop{\rm Re}\nolimits} \left\{ {\bf{x}} \right\}}\\
{{\mathop{\rm Im}\nolimits} \left\{ {\bf{x}} \right\}}
\end{array}} \right].\]
In this work, we focus on the real-valued deep neural network, whose weights structure should be accommodated to complex-valued data.
Let
\[{{\bf{W}}_i} = \left[ {\begin{array}{*{20}{c}}
{{\bf{W}}_i^{\left( 1 \right)}}&{{\bf{W}}_i^{\left( 2 \right)}}\\
{{\bf{W}}_i^{\left( 3 \right)}}&{{\bf{W}}_i^{\left( 4 \right)}}
\end{array}} \right]\]
be the weight matrix in $i$-th layer. In the training stage, we set
 ${{\bf{W}}_i^{\left( 4 \right)}} = {{\bf{W}}_i^{\left( 1 \right)}}$ and ${{\bf{W}}_i^{\left( 3 \right)}} = - {{\bf{W}}_i^{\left( 2 \right)}}$,
which means only half of the weight matrix will be updated. 
This simple operation not only makes the proposed network suitable for
complex-valued OFDM symbols training but also helps to reduce computational complexity.

\subsubsection{Activation Function Design}

\begin{figure}[htbp]
\centerline{\includegraphics[width=0.475\textwidth]{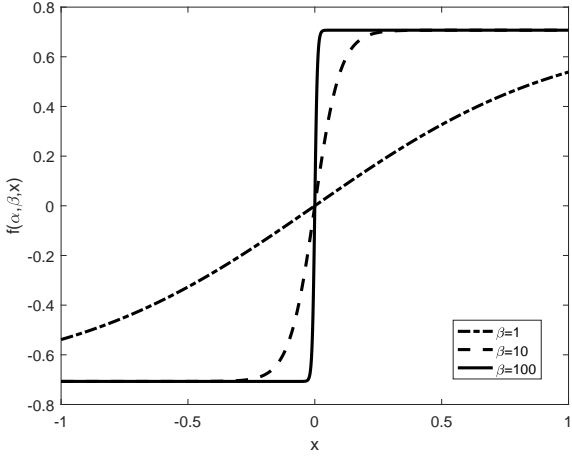}}
\caption{Plot of the proposed functions $f\left( {\alpha ,\beta ,x} \right)$ for $\alpha = 1/\sqrt{2}$, $\beta = 1,10,100$.}
\label{fig_ac}
\end{figure}

As shown in \eqref{eq9}, the data equalization problem itself is a non-convex optimization problem. Discarding the nonconvex constraint yields
the suboptimal solution. In this work, we design a new activation function to help us fully exploit the stochastic gradient descent
method for the discrete optimization problem. Specially, in the case of QPSK signaling and as shown in Fig. \ref{fig_ac},
we propose to use the activation function:
\begin{equation}
\label{eqacd}
f\left( {\alpha ,\beta ,x} \right) = \alpha \frac{{{e^{\beta x}} - {e^{ - \beta x}}}}{{{e^{\beta x}} + {e^{ - \beta x}}}},
\end{equation}
where $\alpha$ is the magnitude of the transmitted signal. $\beta$ is a trainable parameter that controls the layer's outputs, which
will be sufficiently close to the constellation points. Since the data in the higher-order modulation can be represented by a linear
combination of the QPSK data \cite{Mao2007Semidefinite}, we leave out the design for the higher-order modulation.

\subsubsection{Generalization Enhancement}

In this paper, motivated by the pioneer work \cite{Erhan2010Why}, we propose to improve the representation
of the employed neural network by adding an unsupervised loss to \eqref{eq31}. Specially, the proposed neural network based equalizer tries to
update the neural network parameters ${\Theta_1 = \left\{ {{{\bf{W}}_{U+1}}, \cdots ,{{\bf{W}}_{2U}}} \right\}}$ by minimizing
\begin{equation}
\label{eq32}
\begin{array}{*{20}{c}}
{\mathop {{\rm{argmin}}}\limits_{\Theta } }&{{\cal L}\left( \Theta_0  \right) + \lambda {{\cal L}_1}\left( {{\Theta _1}} \right)}
\end{array},
\end{equation}
where
\[{{\cal L}_1}\left( {{\Theta _1}} \right) = \left\| {{{\bf{X}}^{\left( p \right)}} - {\varphi _{2U}}\left( {{{\bf{W}}_{2U}} \cdots {\varphi _{U + 1}}\left( {{{\bf{W}}_1}{{{\bf{\tilde X}}}^{\left( p \right)}}} \right)} \right)} \right\|_2^2,\]
$\Theta = {{\Theta _0} \cup \Theta_1 }$, and $\lambda$ is the penalty parameter that balances $\mathcal{L}\left(\Theta_0\right)$ and $\mathcal{L}_{1}\left(\Theta_{1}\right)$. We use
an unsupervised loss to promote the generalization of the proposed FDNN equalizer for three reasons:
\begin{itemize}
\item The unsupervised loss in \eqref{eq32} itself is a denoising autoender \cite{Vincent2010Stacked} that may help to eliminate
the noise from the data samples.
\item It has been shown in previous studies that learning multi-task jointly can improve the
 generalization error bounds \cite{Maurer2013Excess}. In our case, we jointly minimize
 $\mathcal{L}\left(\Theta\right)$ and $\mathcal{L}_{0}\left(\Theta_{0}\right)$.
\item A nonlinear autoender \eqref{eq32} have been indeed  found to be helpful for key feature
extraction \cite{Vincent2010Stacked, Lecun2015Deep}. In our case, it helps to represent ${{\bf{X}}^{\left( p \right)}}$
 of high-dimension from ${{\bf{U}}^{\left( p \right)}}$ of
low-dimension (the number of users is much less than the number of received antennas).
\end{itemize}

\subsection{Algorithm Implementation and Discussions}

The network model we use consists of eight layers.  The numbers of neurons in each layer are $2R$, $12U$,$6U$, $2U$, $2R$, $2R$, $2R$, and $2U$, respectively. The batch size e  Every 1$k$ bits of the data are grouped and then inputed into the network for training. The Relu function is used as the activation function except the layer $S$. 

The test complexity of the proposed method involves simple matrix-vector multiplications and has complexity of order $O\left( {40{R^2}U} \right)$.  In a MIMO system with 128 transmit antennas and 8 UEs, the average runtime for the proposed method and MMSE are 0.0715 ms and 0.00922 ms, respectively. The runtime performance demonstrates the feasibility of the practical implementation of the proposed method.

\section{Case Studies}
\label{Sec:4}

In this section, using the Tensorflow platform \cite{Abadi2016TensorFlow},
we evaluate the performance of the proposed equalizers via numerical simulations on a PC
with an Intel Core i7-7700K CPU and two NVIDIA GTX 1080 Ti GPUs.

\subsection{System and Network Parameters Setup}

We consider a QPSK modulated MIMO-OFDM system with 128 BS antennas, simultaneously serving 8 users.
Following the IEEE 802.11a standard, we use 64 subcarriers in an OFDM block. The channel tap length is $\mu = 8$.
It is assumed that the coherence time  is $T = 64$. A typical urban environment
with max delay 16 is considered.

\begin{table}[htbp]
\caption{The network parameters employed in the proposed equalizer. }
\begin{center}
\begin{tabular}{|c|c|c|c|}
\hline
\textbf{}&\multicolumn{3}{|c|}{\textbf{Network Parameters}} \\
\cline{2-4}
\textbf{ } & \textbf{\textit{Layer}}& \textbf{\textit{Activation}}& \textbf{\textit{Weights}} \\
\hline
$L_1$& Input& - &  - \\ 
$L_2$& Full-Connected & relu& $256 \times 96$ GM \\
$L_3$& Full-Connected& relu&  $96 \times 48$ GM  \\
$L_4$& Full-Connected& relu&  $48 \times 16$ GM \\
$L_5$& Full-Connected& relu&  $16 \times 48$ GM   \\
$L_6$& Full-Connected& relu&  $48 \times 96$ GM\\
$L_7$& Full-Connected& relu&  $96 \times 256$ GM  \\
$S$ $^{\mathrm{a}}$ & Full-Connected& $f\left( {\alpha ,\beta ,x} \right)$& $16 \times 16$ IM \\
\hline
\multicolumn{4}{l}{$^{\mathrm{a}}${This layer is connected to layer $L_4$ as shown in Fig. \ref{fig1}. }}
\end{tabular}
\label{tab1}
\end{center}
\end{table}

Tab. \ref{tab1} illustrates the parameters in the proposed network
 i.e., the layer size, layer initializations, and activation functions.
 The notation GM in Tab. \ref{tab1} means the Gaussian matrix whose elements have zero mean and variance 0.01. IM means the identity matrix.
The parameter $\beta$ in the proposed activation function $f\left( {\alpha ,\beta ,x} \right)$ will be updated from 1 to 100 to find
the optimal $\beta$ on every mini-batch training.

\subsection{Effect of the Quantization Level}

\begin{figure}[htbp]
\centerline{\includegraphics[width=0.475\textwidth]{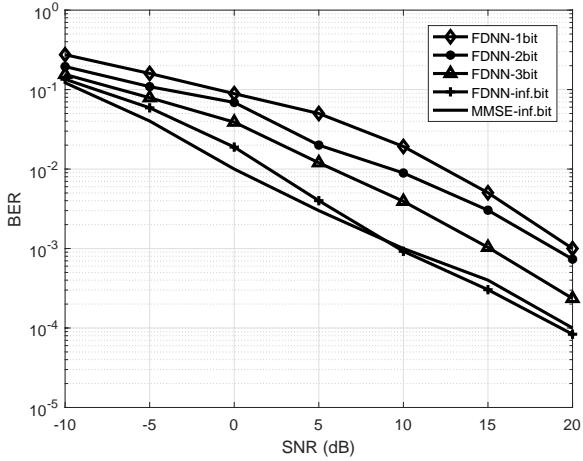}}
\caption{BER curves of the proposed equalizer with different quantization levels. }
\label{fig2}
\end{figure}

In Fig. \ref{fig2}, we investigate the BER performance of the proposed equalizer with different quantization levels.
The benchmark equalizer considered is the MMSE equalizer with infinite-resolution ADCs (MMSE-inf.bit).
It can be observed that the performance of proposed FDNN equalizer improve as a result of a increased bits of ADCs. The performance gap between
the MMSE-inf.bit equalizer and the proposed equalizer is negligible when the number of the bits of ADCs is no less than 2.
Interestingly, the proposed FDNN equalizer with infinite-resolution ADCs (FDNN-inf.bit) slightly outperforms the MMSE-inf.bit equalizer
when the SNR is over 10 dB.
These results above
demonstrate that the proposed network can nicely approximate the nonlinearity introduced by the employed ADCs.

\subsection{The Effect on the Distribution of the Channel Taps}

\begin{figure}[htbp]
\centerline{\includegraphics[width=0.475\textwidth]{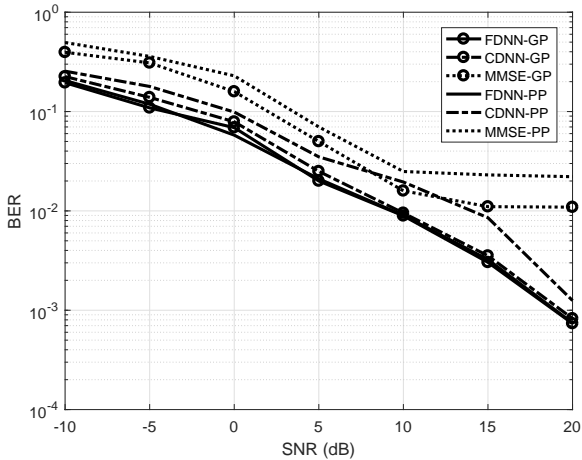}}
\caption{BER curves of the compared equalizers with two kinds of channel taps. The performance is evaluated in a
MIMO system with 2-bit ADCs. The number of taps is 24. }
\label{fig3}
\end{figure}

In Fig. \ref{fig3}, we study the performance of the proposed equalizers (CDNN and FDNN) under two kinds of pilots:
Gaussian distributed pilots (GP) and Poisson distributed
pilots (PP)\footnote{For the case of GP, the taps are assumed be uncorrelated zero-mean Gaussion variables with unit variance.
For the Poisson channel model, a widely used model in the molecular and optical communication systems
\cite{Farsad2018Neural}, the poisson parameter is set to $\lambda = 0.5$.} .
Fig. \ref{fig3} shows BER curves of the MMSE equalizer do not decease distinctly as the SNR increases when the SNR
is above 10 dB. On the other hand, the CDNN and FDNN equalizers have robust performance in a wide range of SNR.
Besides, it is seen that the proposed FDNN equalizer outperforms the CDNN and MMSE equalizer under scenarios where GD pilots and PD pilots
are used.  The FDNN equalizer is more robust than the CDNN equalizer in the case of PD pilots. Such an phenomena may be explained by the enhanced generalization of the FDNN.

\subsection{The Effect of Pilots Numbers}

\begin{figure}[htbp]
\centerline{\includegraphics[width=0.475\textwidth]{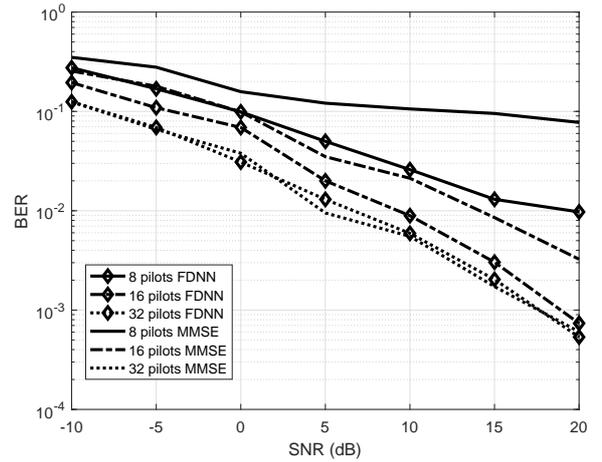}}
\caption{BER curves of the compared equalizers with different lengths of Gaussian pilots. The performance is evaluated in a
MIMO system with 2-bit ADCs.}
\label{fig4}
\end{figure}

The feedbacks (in this paper, we only consider the number of the pilots) are limited in a real communication system,
which reveals another challenge in the design of
an effective neural network. In Fig. \ref{fig4}, we show the results of the effect of the pilot numbers.

From Fig. \ref{fig4}, the larger numbers of pilots lead to better performance of all compared equalizers
and they have comparable performance when 32 pilots are used.
In the case that when only 8 pilots are used, the proposed FDNN equalizer significantly outperforms the MMSE
equalizer. The proposed FDNN equalizer and the MMSE equalizer can respectively achieve a target BER of $10^{-3}$
with 8 pilots and 16 pilots, demonstrating the potential for pilots saving.

\section{Conclusions and Future Work}
\label{Sec:5}

In this paper, the deep neural network based equalizer has been proposed for the MIMO-OFDM systems with low-resolution ADCs.
The experimental results show that the proposed equalizer is robust to different channel taps (i.e., Gaussian, and Poisson) and
significantly outperforms the linearized MMSE equalizer. In addition, given a target of bit error rate, the proposed learning
architecture with an unsupervised loss is more efficient in terms of the number of pilots,
when compared to the learning architecture without such a design or the linearized MSSE equalizer. A key feature of the proposed method is that the employment of all unlabeled data can enhance the quality of the equalization result over different kinds of channel models. 

It will be a long way to go regarding the neural networks based approach as a general way for the MIMO equalization. We shall consider the impact of the change of system parameters and integration into full-duplex communication systems. Nonetheless, it is believed that the proposed method has the potential to impact the design of future communication systems. In our future work, we will provide theoretical analysis, i.e., based on the Rademacher complexity analysis \cite{Bartlett2002R}, for the generalization improvement in the fine deep neural network based equalizer. Besides, it is interesting to investigate how many parameters are needed for training a neural network based method for data equalization problem in
MIMO-OFDM systems with low-resolution ADCs.

\section{Acknowledgment}
\label{Sec:6}

The authors would like to thank Dr. Yanwen Fan, from the EECS department of the University of Tennessee, Knoxville,
for fruitful discussions in the neural network algorithm implementation.

\bibliographystyle{IEEEtran}
\bibliography{leochu-bib}

\begin{thebibliography}{10}
\providecommand{\url}[1]{#1}
\csname url@samestyle\endcsname
\providecommand{\newblock}{\relax}
\providecommand{\bibinfo}[2]{#2}
\providecommand{\BIBentrySTDinterwordspacing}{\spaceskip=0pt\relax}
\providecommand{\BIBentryALTinterwordstretchfactor}{4}
\providecommand{\BIBentryALTinterwordspacing}{\spaceskip=\fontdimen2\font plus
\BIBentryALTinterwordstretchfactor\fontdimen3\font minus
  \fontdimen4\font\relax}
\providecommand{\BIBforeignlanguage}[2]{{%
\expandafter\ifx\csname l@#1\endcsname\relax
\typeout{** WARNING: IEEEtran.bst: No hyphenation pattern has been}%
\typeout{** loaded for the language `#1'. Using the pattern for}%
\typeout{** the default language instead.}%
\else
\language=\csname l@#1\endcsname
\fi
#2}}
\providecommand{\BIBdecl}{\relax}
\BIBdecl

\bibitem{Lu2014An}
L.~Lu, G.~Y. Li, A.~L. Swindlehurst, A.~Ashikhmin, and R.~Zhang, ``An overview
  of massive {MIMO}: {B}enefits and challenges,'' \emph{IEEE J. Sel. Topics
  Signal Process.}, vol.~8, no.~5, pp. 742--758, 2014.

\bibitem{Rusek2012Scaling}
F.~Rusek, D.~Persson, B.~K. Lau, and E.~G. Larsson, ``Scaling up {MIMO}:
  {O}pportunities and challenges with very large arrays,'' \emph{{IEEE} Signal
  Process. Mag.}, vol.~30, no.~1, pp. 40--60, 2012.

\bibitem{Wang2015Multiuser}
S.~Wang, Y.~Li, and W.~Jing, ``Multiuser detection in massive spatial
  modulation {MIMO} with low-resolution {ADCs},'' \emph{{IEEE} Trans. Wireless
  Commun.}, vol.~14, no.~4, pp. 2156--2168, 2015.

\bibitem{Wang2016Signal}
S.~Wang and Z.~Lin, ``Signal processing in massive {MIMO} with iq imbalances
  and low-resolution {ADCs},'' \emph{{IEEE} Trans. Wireless Commun.}, vol.~PP,
  no.~99, pp. 1--1, 2016.

\bibitem{Wang2017Quantization}
F.~Wang, J.~Fang, H.~Li, and S.~Li, ``Quantization design and channel
  estimation for massive {MIMO} systems with one-bit {ADCs},'' \emph{{IEEE}
  Trans. Veh. Technol.}, vol.~PP, no.~99, 2017.

\bibitem{Chu2019E}
L.~Chu, F.~Wen, L.~Li, and R.~C. {Qiu}, ``Efficient nonlinear precoding for
  massive {MIMO} downlink systems with 1-bit {DAC}s,'' \emph{{IEEE} Trans.
  Wireless Commun.}, vol.~18, no.~9, pp. 4213--4224, 2019.

\bibitem{Jacobsson2017Throughput}
S.~Jacobsson, G.~Durisi, M.~Coldrey, U.~Gustavsson, and C.~Studer, ``Throughput
  analysis of massive {MIMO} uplink with low-resolution {ADCs},'' \emph{{IEEE}
  Trans. Wireless Commun.}, vol.~16, no.~6, pp. 4038--4051, 2017.

\bibitem{Mezghani2017Blind}
A.~Mezghani and A.~L. Swindlehurst, ``Blind estimation of sparse broadband
  massive {MIMO} channels with ideal and one-bit {ADCs},'' \emph{{IEEE} Trans.
  Signal Process.}, vol.~PP, no.~99, pp. 1--1, 2017.

\bibitem{Xu2018Performance}
J.~Xu, W.~Xu, H.~Zhang, G.~Y. Li, and X.~You, ``Performance analysis of
  multi-cell millimeter wave massive {MIMO} networks with low-precision
  {ADCs},'' \emph{{IEEE} Trans. Commun.}, pp. 1--1, 2018.

\bibitem{Julian1952Crosscorrelation}
J.~J. Bussgang, ``Crosscorrelation functions of amplitude-distorted {Gaussian}
  signals,'' \emph{MIT Res. Lab. Electron.}, vol. 216, 1952.

\bibitem{jeon2018one}
Y.-S. {Jeon}, N.~{Lee}, S.-N. {Hong}, and R.~W. {Heath}, ``One-bit sphere
  decoding for uplink massive {MIMO} systems with one-bit {ADCs},''
  \emph{{IEEE} Trans. Commun.}, vol.~17, no.~7, pp. 4509--4521, 2018.

\bibitem{myers2019message}
N.~J. {Myers} and R.~W. {Heath}, ``Message passing-based joint cfo and channel
  estimation in millimeter wave systems with one-bit {ADCs}.'' \emph{{IEEE}
  Trans. Commun.}, pp. 1--14, 2019.

\bibitem{Wen2016Bayes}
C.~K. Wen, C.~J. Wang, J.~Shi, K.~K. Wong, and P.~Ting, ``Bayes-optimal joint
  channel-and-data estimation for massive {MIMO} with low-precision {ADCs},''
  \emph{{IEEE} Trans. Signal Process.}, vol.~64, no.~10, pp. 2541--2556, 2016.

\bibitem{Zhang2018On}
J.~Zhang, L.~Dai, L.~Xu, L.~Ying, and L.~Hanzo, ``On low-resolution {ADCs} in
  practical 5g millimeter-wave massive {MIMO} systems,'' \emph{IEEE
  Communications Magazine}, vol.~PP, no.~99, pp. 2--8, 2018.

\bibitem{Lecun2015Deep}
Y.~Lecun, Y.~Bengio, and G.~Hinton, ``Deep learning.'' \emph{Nature}, vol. 521,
  no. 7553, p. 436, 2015.

\bibitem{Ye2018Power}
H.~Ye, G.~Y. Li, and B.~H. Juang, ``Power of deep learning for channel
  estimation and signal detection in {OFDM} systems,'' \emph{{IEEE} Wireless
  Commun. Lett.}, vol.~7, no.~1, pp. 114--117, 2018.

\bibitem{8646357}
H.~{He}, C.~{Wen}, S.~{Jin}, and G.~Y. {Li}, ``A model-driven deep learning
  network for {MIMO} detection,'' in \emph{IEEE GlobalSIP}, 2018, pp. 584--588.

\bibitem{Farsad2018Neural}
N.~Farsad and A.~Goldsmith, ``Neural network detection of data sequences in
  communication systems,'' \emph{{IEEE} Trans. Signal Process.}, vol.~66,
  no.~21, pp. 5663--5678, 2018.

\bibitem{Jeon2016Supervised}
Y.~S. Jeon, S.~N. Hong, and N.~Lee, ``Supervised-learning-aided communication
  framework for {MIMO} systems with low-resolution {ADCs},'' \emph{{IEEE}
  Trans. Veh. Technol.}, vol.~PP, no.~99, pp. 1--1, 2018.

\bibitem{Goldsmith2005Wireless}
A.~Goldsmith, \emph{Wireless Communications}.\hskip 1em plus 0.5em minus
  0.4em\relax New York, NY, USA: Cambridge University Press, 2005.

\bibitem{Chu2019R}
L.~{Chu}, F.~{Wen}, and R.~C. {Qiu}, ``Eigen-inference precoding for coarsely
  quantized massive {MU}-{MIMO} system with imperfect {CSI},'' \emph{{IEEE}
  Trans. Veh. Technol.}, vol.~68, no.~9, pp. 8729--8743, 2019.

\bibitem{Poor1994An}
H.~V. Poor, \emph{An Introduction to Signal Detection and Estimation (2Nd
  Ed.)}.\hskip 1em plus 0.5em minus 0.4em\relax Berlin, Heidelberg:
  Springer-Verlag, 1994.

\bibitem{Zhang17U}
\BIBentryALTinterwordspacing
C.~Zhang, S.~Bengio, M.~Hardt, B.~Recht, and O.~Vinyals, ``Understanding deep
  learning requires rethinking generalization,'' \emph{CoRR}, vol.
  abs/1611.03530, 2017. [Online]. Available:
  \url{http://arxiv.org/abs/1611.03530}
\BIBentrySTDinterwordspacing

\bibitem{NIPS2018_7339}
B.~Hanin, ``Which neural net architectures give rise to exploding and vanishing
  gradients?'' in \emph{NeurIPS}, S.~Bengio, H.~Wallach, H.~Larochelle,
  K.~Grauman, N.~Cesa-Bianchi, and R.~Garnett, Eds., 2018, pp. 582--591.

\bibitem{Mao2007Semidefinite}
Z.~Mao, X.~Wang, and X.~Wang, ``Semidefinite programming relaxation approach
  for multiuser detection of qam signals,'' \emph{{IEEE} Trans. Wireless
  Commun.}, vol.~6, no.~12, pp. 4275--4279, 2007.

\bibitem{Erhan2010Why}
D.~Erhan, Y.~Bengio, A.~C. Courville, P.~A. Manzagol, and S.~Bengio, ``Why does
  unsupervised pre-training help deep learning?'' \emph{J. Mach. Learn. Res.},
  vol.~11, no.~3, pp. 625--660, 2010.

\bibitem{Vincent2010Stacked}
P.~Vincent, H.~Larochelle, I.~Lajoie, Y.~Bengio, and P.~A. Manzagol, ``Stacked
  denoising autoencoders: Learning useful representations in a deep network
  with a local denoising criterion,'' \emph{J. Mach. Learn. Res.}, vol.~11,
  no.~12, pp. 3371--3408, 2010.

\bibitem{Maurer2013Excess}
A.~Maurer and M.~Pontil, ``Excess risk bounds for multitask learning with trace
  norm regularization,'' \emph{J. Mach. Learn. Res.}, vol.~30, pp. 55--76,
  2013.

\bibitem{Abadi2016TensorFlow}
M.~Abadi, P.~Barham, J.~Chen, Z.~Chen, A.~Davis, J.~Dean, M.~Devin,
  S.~Ghemawat, G.~Irving, and M.~Isard, ``Tensorflow: a system for large-scale
  machine learning,'' in \emph{12th {USENIX} Symposium on Operating Systems
  Design and Implementation}, 2016, pp. 265--283.

\bibitem{Bartlett2002R}
P.~Bartlett and S.~Mendelson, ``Rademacher and gaussian complexities: Risk
  bounds and structural results,'' \emph{J. Mach. Learn. Res.}, vol. Nov.,
  no.~3, pp. 224--240, 2002.

\end{thebibliography}

\end{document}